\documentstyle[preprint,eqsecnum,aps]{revtex}
\tightenlines

\newcommand{\cs}{\'{c}}
\newcommand{\ch}{\v{c}}

\newcommand{\beq}{\begin{equation}}
\newcommand{\eeq}{\end{equation}}
\newcommand{\bdm}{\begin{displaymath}}
\newcommand{\edm}{\end{displaymath}}
\newcommand{\beqa}{\begin{eqnarray}}
\newcommand{\eeqa}{\end{eqnarray}}
\newcommand{\beqab}{\begin{eqnarray*}}
\newcommand{\eeqab}{\end{eqnarray*}}

 \def\@makefnmark{\hbox to 0pt{$^{\@thefnmark}$\hss}}  

\newcounter{saveeqn}%

\begin{document}
\draft
\preprint{}
\title{Violation of the string hypothesis and Heisenberg
 XXZ spin chain}  
\author{Amon Ilakovac\footnote{e-mail: ailakov@phy.hr},
Marko Kolanovi\cs\footnote{e-mail: kmarko@phy.hr},
Silvio Pallua\footnote{e-mail: pallua@phy.hr} and
Predrag Prester\footnote{e-mail: pprester@phy.hr} }
\address{Department of Theoretical Physics, University of Zagreb\\
Bijeni\ch ka c.32, POB 162, 10001 Zagreb, Croatia}
\date{\today}
\maketitle
\begin{abstract}
In this paper we count the numbers of real and complex solutions to Bethe
constraints in the two particle sector of the XXZ model. We find exact number
of exceptions to the string conjecture and total number of solutions 
which is required for completeness.
\end{abstract}

\narrowtext

\section{Introduction}

Integrable spin chains have proven to be useful in studying various theoretical
ideas in field theory and statistical physics. In continuum limit, one can relate
spin chains to massive Thirring model, sine-Gordon theory, Liouville theory and
others \cite{Luth}, \cite{ft}. Faddeev and
Korchensky \cite{Fadd} suggested their possible relevance for QCD. Connection to 
matrix models was also suggested \cite{Matr}. A very successful method in solving
spin chains and in general integrable models both on a lattice and in the continuum is the 
 Bethe ansatz \cite{Beth,Orba}. Despite the fact that a lot is known
about this method there is still one set of open questions  concerning the so called
string conjecture \cite{Taka,Gaud}.

The Bethe ansatz method leads to a set of transcendental equations (called Bethe 
constraints) for momenta of quasiparticles. In usual search for solutions of these 
equations a simplifying assumption is made, the already mentioned string conjecture. 
This conjecture, which we shall afterwards formulate more precisely, classifies the 
complex solutions for momenta of quasiparticles. It is well known that there are 
exceptions to the string conjecture near the antiferromagnetic ground state
\cite{Woyn,Babe,Dest}. Recently, exceptions have been found \cite{Essl,Isle}
already in two particle sector of the XXX spin chain. Similar results have been
found for the Hubbard model \cite{hb1}. In the case of the XXX spin chain the number of missing 
solutions (compared to the string conjecture prediction) was found to be $\sqrt{N}$
where $N$ is the number of degrees of freedom. A certain class of real solutions not 
allowed by the string conjecture was observed by J\"{u}ttner and D\"{o}rfel 
\cite{Jutt} in the XXZ chain. However, a systematic investigation of complex solutions
and thus of exceptions to the string conjecture is missing.

There are several reasons why it would be desirable to understand the limits of
validity of the string conjecture or equivalently to have a clear understanding of 
nature and number of real and complex solutions for momenta of quasiparticles. One 
reason is that it was used in literature as a tool to obtain various results. One 
example are for instance completeness proofs of Bethe states \cite{Beth,hb1,hb,Kiri,Fore,Karo,Fore1}.
Other example are investigations which use a lattice regularization of field theoretical
models \cite{Fenl}. In such cases the results at orders which are lower than $N$
may depend on modifications of even a single root as these authors stress. As is well
known the string conjecture was also used by Bergknoff and Thacker \cite{Thac}
in deriving breather states of the massive Thirring model. This was recently criticised
on the basis of numerical analysis of Bethe equations. A recent numerical calculations 
independent of Bethe ansatz and based on the lattice regularization \cite{Thir} led to
the usual bound state spectrum of the massive Thirring model thus suggesting that the question
raised by previously mentioned authors could be related to understanding of the
string conjecture and its violations. This result is based on assumption of equivalence 
of sine-Gordon and the massive Thirring model. Another calculation for the massive Thirring
model itself is in progress \cite{thir1}.

In this paper we shall classify all solutions (both complex and real) in two particle 
sector of the XXZ model. That will allow us to find the number of exceptions to the string
conjecture for a given coupling constant and a given number of lattice sites $N$. We shall
in particular find that the number of exceptions to the string conjecture in thermodynamical limit
is finite, except for the value of the coupling constant in which it coincides with the XXX
model and what is consistent with previously found result \cite{Essl,Isle}.

We shall consider the XXZ spin chain defined with the following Hamiltonian
\beq   \label{ham}
H=-\frac{1}{2}\sum_{n=1}^{N}\left( \sigma_{n}^{x}\sigma_{n+1}^{x}+\sigma_{n}^{y}
\sigma_{n+1}^{y}+\Delta\sigma_{n}^{z}\sigma_{n+1}^{z}  \right)
,\quad\vec{\sigma}_{N+1}\equiv\vec{\sigma}_{1}.
\eeq

This Hamiltonian acts in $N^{2}$ dimensional Hilbert space  $\mathcal{H}=(\otimes  C^2)^N$
. In the Bethe ansatz method one introduces 
the basis states $|n_{1}...n_{M}\rangle$ with $M$ spins down, where the numbers 
$n_{1}, ..., n_{M}$ denote the lattice positions of the down spins. With $|0\rangle$ we
denote the state with all spins up. A general element of the above defined Hilbert space
and thus in particular the eigenstates of the Hamiltonian can than be written in the sector 
with $M$ spins down as  
\beq \label{psi}
|\psi_{M}\rangle=\sum_{1\leq n_{1}\leq n_{2}\leq ...\leq n_{M}\leq N}  \psi_{M}
(n_{1}...n_{M})|n_{1}...n_{M}\rangle.
\eeq
The Bethe method consists in searching for Hamiltonian eigenstates in the form
\beq \label{psi1}
\psi_{M}(n_{1}...n_{M})=\sum_{P}\exp\left\{i(\sum_{j=1}^{M}k_{P_{j}}n_{j}+
\frac{1}{2}\sum_{1\leq j\leq l\leq M} \phi_{Pj,Pl})\right\},
\eeq
where the sum runs over elements of the permutation group $S_{M}$. The momenta
$k_{i},\ i=1,..., M$ and phase shifts $\phi_{j,i}$ have to be determined from the eigenvalue 
equation and periodicity requirement on functions  $\psi_{M}(n_{1}...n_{M})$. The well
known procedure gives following expressions for phase shift  $\phi_{j,i}$, energy $E$ and
 momentum $P$ in terms of pseudomomenta  $k_{i},\ i=1,..., M$  
\beq \label{phi}
 \phi_{j,i}=2\arctan \frac{\Delta\sin\frac{(k_{j}-k_{i})}{2}}{\cos\frac{(k_{j}+k_{i})}{2}
 - \Delta\cos \frac{(k_{j}-k_{i})}{2}},
\eeq
\beq \label{e}
E=-\frac{N\Delta}{2}+2\sum_{i=1}^{M}(\Delta-\cos k_{i}),
\eeq
\beq \label{p}
P=\sum_{i=1}^{M}k_{i}.
\eeq
The periodicity requirement leads to the following constraints for the momenta of quasiparticles
\beq \label{bet}
Nk_{i}+\sum_{j=1}^{M}\phi_{i,j}=2\pi\lambda_{i},\quad i=1, ..., M.
\eeq
The $M$ Bethe numbers $\lambda_{i},\ i=1, ..., M$ are half integers (integers) for $M$ even (odd).
Thus for $M$ even we can chose $\lambda_{i}\in\{-\frac{N-1}{2},..., \frac{N-1}{2}\}$ for $N$
even and  $\lambda_{i}\in\{-\frac{N}{2}, ..., \frac{N}{2}-1\}$ for $N$ odd. Sometimes it is 
useful to introduce a transformation from pseudomomenta $k_{i},\ i=1,..., M$ to rapidity
variables $x_{i},\ i=1,..., M$  with following relation
\beq \label{rap}
\cot\frac{k_{i}}{2}=\cot\frac{\theta}{2}\tanh\frac{\theta x_{i}}{2},\quad\Delta=\cos\theta.
\eeq
In this parametrisation Bethe constraints read
\beq  \label{pro}
\left\{\frac{\sinh \frac{\theta}{2}(x_{k}+i)}{\sinh\frac{\theta}{2}(x_{k}-i)}\right\}^N=
-\prod_{l=1}^{M}\left\{\frac{\sinh \frac{\theta}{2}(x_{k}-x_{l}+2i)}
{\sinh \frac{\theta}{2}(x_{k}-x_{l}-2i)}\right\},\quad k=1, ..., M.
\eeq
The string conjecture states that solutions of these equations form string configurations 
with rapidities that are forming strings of length $n$. Rapidities in string have common
real parts and equidistant imaginary parts. More precisely, a string of order
(length) $n$
and parity $+$ or $-$ is a set of $n$ rapidities 
\beqa \label{str}
x_{a,+}^{n,k}&=&x_{a}^{n}+(n+1-2k)i+O(\exp(-\delta N))\quad(\bmod \frac{2\pi}{\theta})\\
x_{a,-}^{n,k}&=&x_{a}^{n}+(n+1-2k)i+\frac{i\pi}{\theta}+O(\exp(-\delta N))\quad(\bmod \frac{2\pi}{\theta})
\eeqa
where $\delta\ge 0,\ k=1, ..., n$ and $x_{a}^{n}$ is real. Insertion of these assumed forms in
(\ref{pro}) gives equations for real parts of strings, which are similar to (\ref{bet}) with
one common Bethe number (integer) $I$ for each string. In addition, a part of the string conjecture was that no 
two strings of the same length can have same integers $I$. This assumptions together with inequalities derived
in \cite{Taka} for numbers $I$ allow one to count the number of string solutions of equations 
(\ref{pro}). In this paper we shall not use equations which are consequence of string conjecture.
However, for the future comparison we mention that in the sector $M=2$ the following number
of solutions for strings of length 2 can be obtained
\beq \label{nst}
N_{S}=2[\frac{1}{2\pi}(N-4)(\pi-2\theta)]+1
\eeq
where $[x]$ denotes integer part of x.
  
\section{Two particle sector and complex solutions}
\setcounter{equation}{0}

We want now to analyse Bethe equations without assuming the string conjecture. For 
simplicity we shall treat the two particle sector. In this sector Bethe constraints
(\ref{bet}) read
\beqa\label{tps}
Nk_{1}+\phi_{1,2}=2\pi\lambda_{1},\\
\label{tps1} Nk_{2}-\phi_{1,2}=2\pi\lambda_{2}.
\eeqa
Here we want in particular to look for complex solutions. Due to reality of energy
and momentum, $k_{1}$ and $k_{2}$ have to be complex conjugates of each other
\beqa \label{k1k2}
k_{1}=k_{r}+ik_{i},\\
k_{2}=k_{r}-ik_{i}.
\eeqa
We can express $k_{r}$ and $k_{i}$ by taking sum and difference of equations
(\ref{tps}) and (\ref{tps1})
\beq \label{kr}
k_{r}=\frac{\pi}{N}(\lambda_{1}+\lambda_{2}),
\eeq
\beq \label{ki}
iNk_{i}=\pi(\lambda_{1}-\lambda_{2})-2\arctan\frac{\Delta\sin(ik_{i})}
{\cos k_{r}-\Delta\cos(ik_{i})}.
\eeq
Further straightforward manipulation allows to introduce a simple equation
for $k_{i}$ in terms of $k_{r}$. So the final set of equations, which we shall
consider, is the equation (\ref{kr}) for $k_{r}$ and the equations for $k_{i}$
\beq \label{sh}
\frac{\sinh (k_{i}(\frac{N}{2}-1))}{\sinh (k_{i}\frac{N}{2})}=\frac{\cos k_{r}}{\cos\theta}
\qquad\lambda_{1}+\lambda_{2}\quad \mbox{odd},
\eeq
\beq \label{ch}
\frac{\cosh (k_{i}(\frac{N}{2}-1))}{\cosh (k_{i}\frac{N}{2})}=\frac{\cos k_{r}}{\cos\theta}
\qquad\lambda_{1}+\lambda_{2}\quad \mbox{even}.
\eeq
We shall distinguish solutions of equation (\ref{sh}) and call them s-solutions (strings)  from
those of equation (\ref{ch}) which we shall call c-solutions (strings). In fact these equations 
will give a basis for a natural classification of solutions. Any solution in the two particle sector
will depend only on the sum of Bethe numbers and its parity. A choice of different Bethe numbers
that gives the same sum (e.g. $\left(\frac{3}{2},-\frac{1}{2}\right),\left(\frac{1}{2},\frac
{1}{2}\right))$ corresponds to taking different branches of the phase shift in (\ref{tps}) and
(\ref{tps1}).  As we shall see later,
number of solutions will be different in these two classes and exceptions to the string conjecture will 
be due to the class s only. As the sum of Bethe numbers can be taken between $-N+1$ and $N-1$ for $N$ even and
between $-N$ and $N-2$ for $N$ odd, we see that $k_{r}$ can take $2N-1$ different equidistant values 
between $-\pi$ and $\pi$. From equations (\ref{sh}) and (\ref{ch}) (and Fig. 1) we see that
admissible interval for $\cos k_{r}$ is
\beqa \label{int1}
\mbox{s-strings}&:&\quad 0\leq\cos k_{r}<\Delta(1-\frac{2}{N}),
\\ \mbox{c-strings}&:&\quad 0\leq\cos k_{r}<\Delta
\eeqa  
for $\Delta\ge 0\quad (0\leq\theta\le\frac{\pi}{2})$ and
\beqa \label{int2}
\mbox{s-strings}&:&\quad \Delta(1-\frac{2}{N})<\cos k_{r}\leq 0,
\\ \mbox{c-strings}&:&\quad \Delta<\cos k_{r}\leq 0
\eeqa
for $\Delta\le 0 \quad(\frac{\pi}{2}\leq\theta\le\pi)$. The energy of complex solutions, according to
 (\ref{e}), will be given with
\beq \label{e1}
E=4\Delta-4\cos k_{r}\cosh k_{i}.
\eeq
Here we measure the energy from the referent state with all spins up. Due to relations 
(\ref{ch}) and (\ref{sh}) one can see that energy intervals for complex solutions are
\beqa \label{e2}
0< E(\mbox{c-strings})\leq 2\Delta,\\
\frac{8\Delta}{N}< E(\mbox{s-strings})\leq 2\Delta.
\eeqa
Now the left side of both equations
(\ref{sh}), (\ref{ch}) are monotonously decreasing functions so we shall have a solution for
$k_{i}$ for any $k_{r}$ whose $\cos k_{r}$ is in the previously mentioned interval. For large
$N$ we can approximate admissible interval for s-strings with that
for c-strings. In that case complex solutions will exist if their real parts satisfy
inequality $0\leq\cos k_{r}\leq\Delta$. As we have $\frac{2N-1}{2\pi}$ solutions per 
unit $k_{r}$ interval, we conclude that 
\beq \label{nta}
\frac{1}{2\pi}(2N-1)(\pi-2\theta)
\eeq
solutions in form of strings can be obtained. This is consistent (up to at most two solutions) 
with the string conjecture and result (\ref{nst}).   

\section{Number of bound states (complex solutions) and violation of the string hypothesis}
\setcounter{equation}{0}

We want to determine the number of bound states as a function of the coupling constant $\Delta$
and number of sites $N$. We shall first consider complex solutions for fixed $N$ and 
different $\Delta$. In Fig. 2 the case $N=40$ is presented. For each (calculated) 
$\Delta$ real parts of possible complex solutions are given. We see that in region of 
negative coupling constant the complex solutions are present for
$\frac{\pi}{2}\leq|k_{r}|\leq \pi$ and in the region of positive coupling constant 
for $0\leq|k_{r}|\leq \frac{\pi}{2}$. As $k_{r}$ tends to
$\frac{\pi}{2}$, $k_{i}$ increases and so the localisation of two spins down
increases (notice that the ratio of probability amplitudes for finding spins down
 on lattice sites $n_{1}$ and $n_{2}$ is proportional to $\exp -|k_{i}(n_{1}-n_{2}+1)|$).
 As we decrease the coupling constant $|\Delta|$, bound states with $|k_{r}|\ge|\theta|$  
for $\Delta\le 0$ and with $|k_{r}|\le|\theta|$ for $\Delta\ge 0$ disappear. These are
states with the smallest localisation. The bound states with high localisation ($k_{i}\ \mbox{
high},\ k_{r}\approx \frac{\pi}{2}$) exist in almost all the region of coupling constant and
disappear near the free theory point $(\Delta=0)$.
In Fig. 3 and 4 we present numerical analysis of $N$ dependence of string
solutions for $\Delta\neq 1$ and $\Delta=1$. In $\Delta=1$ case c-strings are allowed
for all values of $-\frac{\pi}{2}\le k_{r}\le\frac{\pi}{2}$ and so their number rises
linearly with $N$ as predicted by (\ref{nta}) and the string conjecture. However, the number of
s-strings rises also linearly with $N$ until the real parts do not reach the region where
$1-\frac{2}{N}\le\frac{\cos k_{r}}{\Delta}\le 1$ when such s-string is no more a solution
of Bethe equations. So first two strings disappear for $N=22$, next for 62, 121 etc. 
Simultaneously with disappearance of s-strings two real solutions with the same (odd) sum of
Bethe numbers appear. Odd sum of Bethe numbers can be accomplished by two equal Bethe
numbers (which is found by numerically calculation, which favours the choice of
the principal branch of the phase shift) 
and both properties (disappearance of complex solution and appearance of real 
solution with two same Bethe numbers) represent violations of the string conjecture. These
results are consistent with the results of \cite{Essl,Isle}. For $\Delta\neq 1$ however 
we shall find a different result. From Fig. 3 we can see that again for certain values
of $N$ s-strings will disappear and evolve in two close real momenta for which we find identical Bethe
numbers. These exceptional real solutions will disappear again after some $N$, when followed by
 the numerical iteration method. This is in contrast to the
 $\Delta=1$ case. In fact when the solutions are described by more natural classification (equations
 (\ref{sh}) and (\ref{ch}) for complex solutions and equations (\ref{sin}) and (\ref{cos}) for
  real solutions) one could follow their further development. However, here we were interested
  specifically in the choice of equal Bethe numbers when solving equations (\ref{tps}) and (\ref{tps1})
  directly.
  
   We proceed now to give an analytical expression for
number of exceptions to the string conjecture. Due to the previous discussion we
find that the exceptions arise only due to the equation (\ref{sh}), which has no solutions in
the following interval for the sum of Bethe numbers
\beq \label{int3}
(1-\frac{2}{N})\leq\frac{\cos \frac{\pi}{N}(\lambda_{1}+\lambda_{2})}{\Delta}\leq
1\quad\quad(\lambda_{1}+\lambda_{2})\quad \mbox{odd}.
\eeq
Now consider inequality (\ref{int3}) first for $\Delta=1$. The maximal $k_{r}$ for which s-solutions
would still not be possible can be found by expanding  $\cos\frac{\pi}{N}(\lambda_{1}+
\lambda_{2})$ around zero. We find
\beq \label{uv1}
(\lambda_{1}+\lambda_{2})^2<(\frac{2}{\pi}\sqrt N)^2,
\eeq
where $N$ is number of sites after which two complex solutions (for $+k_{r}$ and
$-k_{r}$) disappear and
become solutions with two real momenta and $\lambda_{1}=\lambda_{2}$. For $\lambda_{1}+\lambda_{2}=
1, 3, 5, 7,...$ we get $N=3, 22, 62, 121, ...$ . As previously said, this is consistent with
\cite{Essl,Isle}. 

Now we   turn to general $\Delta\neq 1$. Consider $k_{r}^1$ and $k_{r}^2$ which are just on the 
edges of the interval (\ref{int3}). They satisfy 
\beq \label{k1k2a}
2\sin(\frac{k_{r}^1+k_{r}^2}{2})\sin(\frac{k_{r}^1-k_{r}^2}{2})=\frac{2\Delta}{N}.
\eeq
From this relation the interval $\delta k_{r}$ for which s-strings are missing is given with
\beq \label{dk}
\delta k_{r}=2\arcsin
\frac{\Delta}{N\sin(\frac{\arccos\Delta+\arccos\Delta(1-\frac{2}{N})}{2})}.
\eeq
Number of s-strings per unit interval of $k_{r}$ is 
\beq \label{ui}
\frac{1}{2}\frac{2N-1}{2\pi}2.
\eeq
Here $\frac{2N-1}{2}$ is due the fact that we have to count the number of odd values of $\lambda_{1}+
\lambda_{2}$. As solutions come in pairs (positive and negative total momenta) we need the last factor
of two. Finally number of missing strings is an integer part of
\beq \label{nnn}
n=\frac{2N-1}{\pi}\arcsin\frac{\Delta}{N\sin(\frac{\arccos\Delta+\arccos\Delta(1-\frac{2}{N})}{2})}.
\eeq
The function (\ref{nnn}) is shown in Fig. 5 for few values of $N$. We can see that the number of
missing strings is finite for $\Delta\neq 1$ and that there is no violation of the string 
hypothesis  below some value of coupling constant $\Delta$.
These strings would have energies (\ref{e1}) in the forbidden interval $0\leq E\leq\frac{8\Delta}{N}$
which is near the energy of the state with all spins up.

\section{Number of real solutions and completeness problem}
\setcounter{equation}{0}

In this chapter we shall search for real solutions of Bethe equations. We start again from
the equations  (\ref{tps}) and (\ref{tps1}). After manipulating their difference and sum 
we obtain following equations
for $k=k_{1}+k_{2}$ and $k_{1}-k_{2}$
\beq \label{pp}
\frac{k}{2}=\frac{\pi}{N}(\lambda_{1}+\lambda_{2})
\eeq
\beq \label{sin}
\frac{\sin (\frac{(k_{1}-k_{2})}{2}(\frac{N}{2}-1))}{\sin (\frac{(k_{1}-k_{2})}{2}\frac{N}{2})}
=\frac{\cos \frac{k}{2}}{\cos\theta}
\qquad\lambda_{1}+\lambda_{2}\quad \mbox{odd}
\eeq
\beq \label{cos}
\frac{\cos (\frac{(k_{1}-k_{2})}{2}(\frac{N}{2}-1))}{\cos (\frac{(k_{1}-k_{2})}{2}\frac{N}{2})}
=\frac{\cos \frac{k}{2}}{\cos\theta}
\qquad\lambda_{1}+\lambda_{2}\quad \mbox{even}.
\eeq
From condition (\ref{pp}) we can find number of different momenta $k$. As was already mentioned
there are $2N-1$ different values of $\lambda_{1}+\lambda_{2}$. But changing 
$\lambda_{1}+\lambda_{2}$ by $N$ is equivalent with changing one quasimomentum by $2\pi$
what gives the same solution. This reduces the number of possible values 
$\lambda_{1}+\lambda_{2}$ to $N$ e.g. $\lambda_{1}+\lambda_{2}=0,1, ...,N-1$. Left hand sides 
of equations (\ref{sin}) and (\ref{cos}) are periodic functions. Thus in principle for each of $N$ different
fixed values of the right hand side one can count number of solutions by counting number of
intersections. For a given value of the $\lambda_{1}+\lambda_{2}$ and $\Delta$ we can find following
 number $X$ of intersections for $N$ even
\beqa \label{nev}
\left|\frac{\cos\frac{k}{2}}{\Delta}\right|>1,\quad X=\left(\frac{N}{2}\right
);\quad \left|\frac{\cos\frac{k}{2}}{\Delta}\right|<1,
\quad X=\left(\frac{N-2}{2}\right)\quad  \\
\left|\frac{\cos\frac{k}{2}}{\Delta}\right|>1-\frac{2}{N},\quad X=\left(\frac{N-2}{2}
\right);\quad 
\left|\frac{\cos\frac{k}{2}}{\Delta}\right|<1-\frac{2}{N},
\quad X=\left(\frac{N-4}{2}\right)\quad
\eeqa
for $\lambda_{1}+\lambda_{2}$ even and odd, respectively.
 For $N$ odd
\beqa \label{nod}
\left|\frac{\cos\frac{k}{2}}{\Delta}\right|>1,\quad X=\left(\frac{N-1}{2}\right)
;\quad \left|\frac{\cos\frac{k}{2}}{\Delta}\right|<1,
\quad X=\left(\frac{N-3}{2}\right)\quad \\
\left|\frac{\cos\frac{k}{2}}{\Delta}\right|>1-\frac{2}{N},\quad X=\left(\frac{N-3}{2}
\right);\quad 
\left|\frac{\cos\frac{k}{2}}{\Delta}\right|<1-\frac{2}{N},
\quad X=\left(\frac{N-3}{2}\right)\quad
\eeqa
for $\lambda_{1}+\lambda_{2}$ even and odd, respectively.
When right hand side of equations (\ref{cos}) and (\ref{sin}) becomes smaller than $1$ 
and $1-\frac{2}{N}$ respectively,
corresponding real solution (in fact a pair with $\pm k$) disappears and we get a pair of complex
solutions with positive and negative real parts. Now we can proceed to obtain full number of solutions.
For $N$ even we have $\frac{N}{2}\quad(\frac{N}{2})$ possible values for $\lambda_{1}+\lambda_{2}$ even
(odd). For $N$ odd there are $\frac{N-1}{2}\quad(\frac{N+1}{2})$ possible values for $\lambda_{1}+
\lambda_{2}$ even (odd). Together with results from previous chapter on complex solutions one can
count total number of real and complex solutions. It is important to realise that disappearance of pair
of real solutions results in formation of a two complex solutions and vice versa. Let us count number
of solutions in two extreme cases $\Delta\rightarrow 0$  and  $\Delta\ge\frac{N}{N-2}$. For $\Delta\rightarrow 0$ there are no complex solutions and the
number of real solutions is $N_{real}=\frac{N^2}{2}-\frac{N}{2}={N \choose 2}.$ 
 For $\Delta\ge\frac{N}{N-2}$ number of complex solutions is $N$ $(\frac{\cos k_{r}}{\Delta}\leq 1-
 \frac{2}{N})$ and from (\ref{nev}) and (\ref{nod}) number of real solutions is $N_{real}=\frac{N^2}{2}-
 \frac{3N}{2}={N \choose 2}-{N \choose 1}$. Again total number of solutions is ${N \choose 2}$. We conclude that we find
 ${N \choose 2}$ solutions of Bethe equations (\ref{tps}) and (\ref{tps1}) for every value of $\Delta$.
  We stress that this result is obtained without
 assuming string conjecture, which is usually assumed in completeness proofs. In fact as we discuss in
 this paper string conjecture has exceptions. However, they do not affect completeness proofs because
 disappearance of complex solution (bound state) results in appearance of real solution and vice versa
 so that total number is unchanged.
 
 As we have explained already, analysis was done for simplicity reasons in the two particle sector. Of
 course a systematic analysis for higher sectors may be desirable but it is much more complicated.
 However, we mention some preliminary results for the $M=3$ sector. By numerical analysis we search for
 exceptions to string conjecture among real solutions with coinciding two or all Bethe numbers. They
 are exhibited in Fig. 6 and Fig. 7 for $\Delta=1$ and $\Delta=0.95$. They show similar
 regularities as $M=2$ case. In particular with appearance of real solutions violating string
 conjecture in two particle sector in the three particle sector such exceptions arise in the
 form of perturbed pair of near momenta (of identical Bethe numbers in $M=2$ sector) and a
 third almost independent momenta with a distinct Bethe number. For instance Fig. 6 for
 $\Delta=1$ shows appearance of such solutions around $N=22$ similar to the Fig. 4 for
 $M=2$ and $\Delta=1$. On the Fig. 7 for $\Delta=0.95$ we see that such solutions are found in finite
 intervals of $N$. This is again same as in $M=2$ case (Fig. 3). Finally this preliminary 
 investigation for larger $M$ rises hope that a simple pattern for exception to string 
 conjecture could arise.

\section{Conclusion}
\setcounter{equation}{0}
f
In this paper we count all complex and real solutions of Bethe equations in the two particle
sector. The complex solutions are classified in two classes. For one of them (s-class) the
sum of Bethe numbers is odd and for the other (c-class) it is even. We are able to count
number of solutions in each class for a given coupling constant $\Delta$ and number $N$ of
lattice sites. In such a way we are able to check the validity of usual string conjecture. We
find that there are exceptions to string conjecture and that they are entirely due to the
s-class of solutions. In particular in the thermodynamic limit we show that number of these exceptions is
finite for $\Delta\neq 1$ contrary to the $\Delta=1$ case where it was previously known that
it is infinite. Finally we also show independently of the string conjecture that number of
all solutions is ${N \choose 2}$ and which is required for completeness. The usual proofs
of completeness rely on string hypothesis. Some preliminary numerical results have been
presented  also for three particle sector. These results suggest that similar pattern
observed in two particle sector persists also for larger sectors.

%


\begin{figure}
\caption{Graphical description of the left hand sides of equations (\ref{sh}) and (\ref{ch})
for some values of $N$.
}
\end{figure}

\begin{figure}
\caption{Real parts of complex solutions for $-1\leq\Delta\leq 1$ and number of sites 
$N=40$. Empty squares denote solutions of equation (\ref{sh}) (s-strings) and full circles 
denote solutions of equation (\ref{ch}) (c-strings).
}

\end{figure}

\begin{figure}
\caption{This figure shows dependence of real part of complex solutions on numbers of sites
$N$ for $\Delta=0.95$. It clearly illustrates transmutation of one complex solution in real solution
(two quasimomenta) for a given critical $N$. These two real quasimomenta correspond to same Bethe 
numbers and are obtained by numerical iteration of equations (\ref{tps}) and (\ref{tps1}).
}

\end{figure}

\begin{figure}
\caption{Same as Fig.~3 but for $\Delta=1$. We see that exceptional real solutions with same Bethe
numbers which appear at some critical $N$ persist for all $N\ge N_{crit.}$.
}

\end{figure}

\begin{figure}
\caption{Number of missing strings is an integer part of real number $n$ which is given as a function
of the coupling constant $\Delta$ for three different values of $N$.
}

\end{figure}

\begin{figure}
\caption{All real solutions (triplets of quasimomenta) in $M=3$ sector with at least two identical
Bethe numbers are given for different number of sites and $\Delta=1$. They are obtained by
numerical iteration of Bethe equations (\ref{bet}). 
}

\end{figure}

\begin{figure}
\caption{Same as Fig.~6 but for $\Delta=0.95$.}

\end{figure}

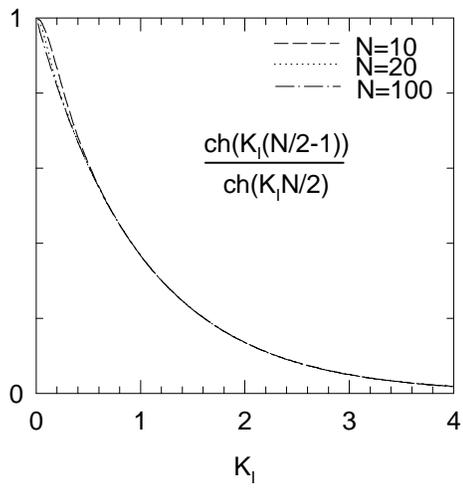
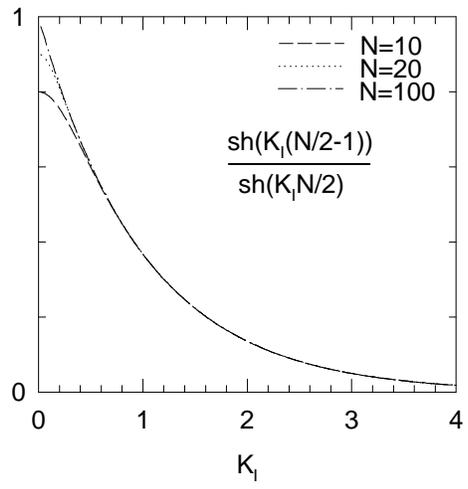
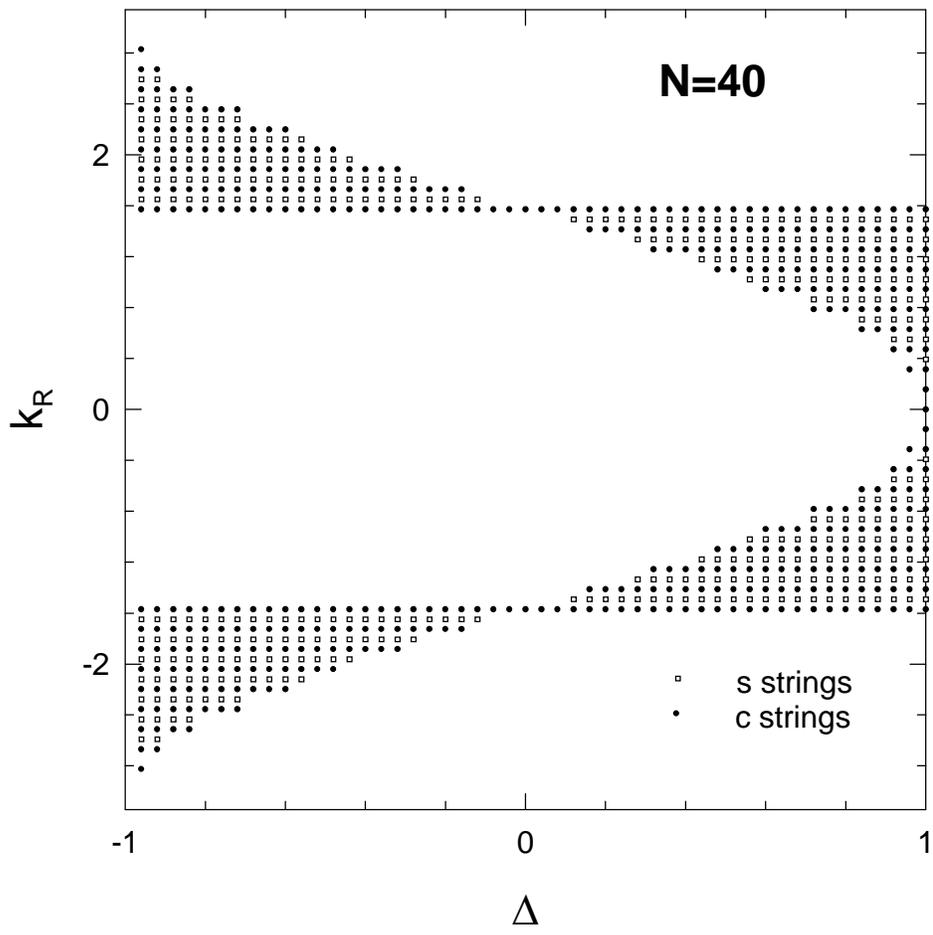
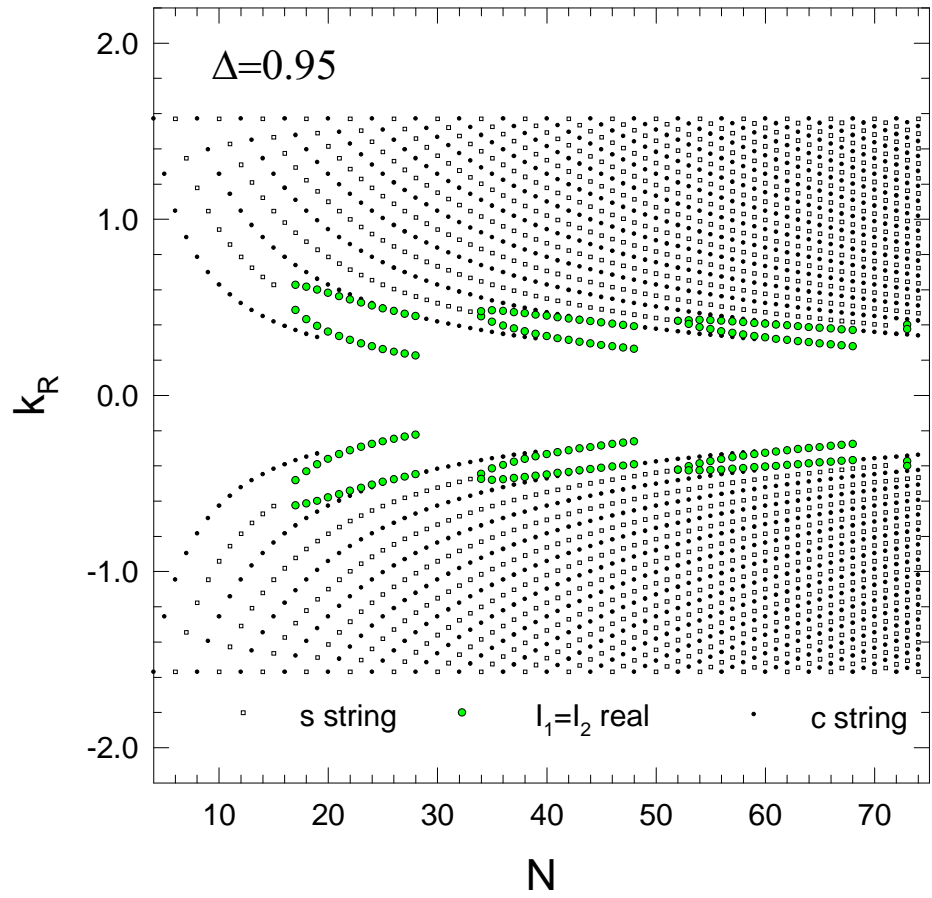
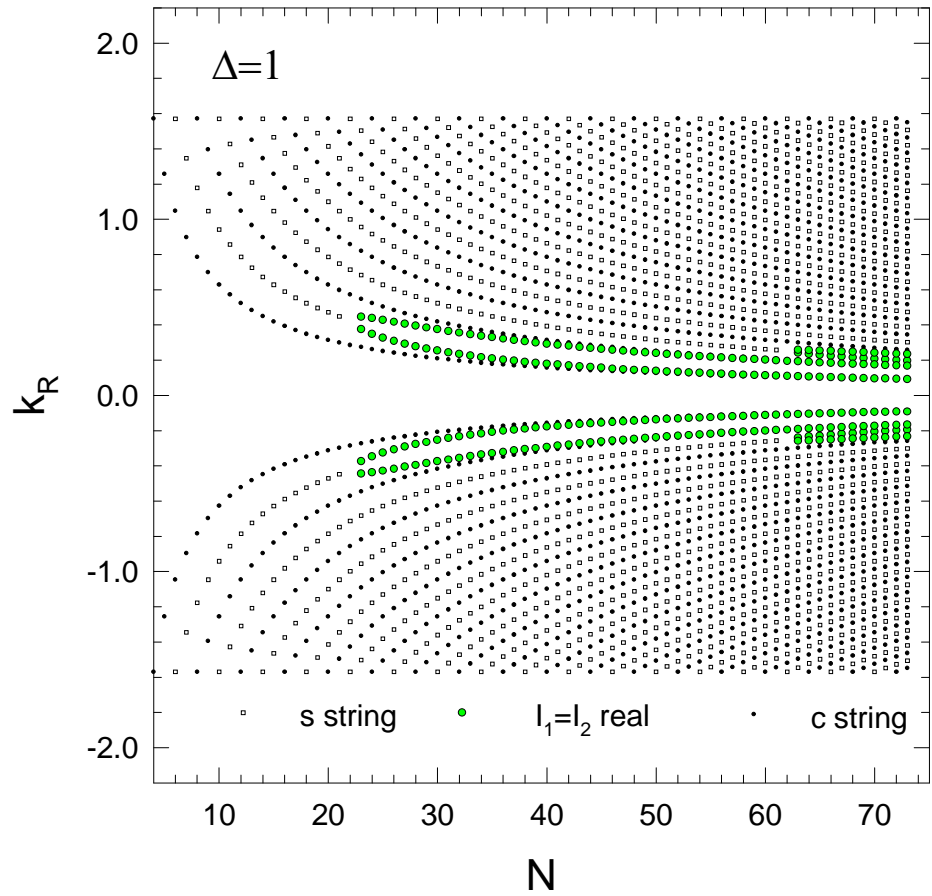
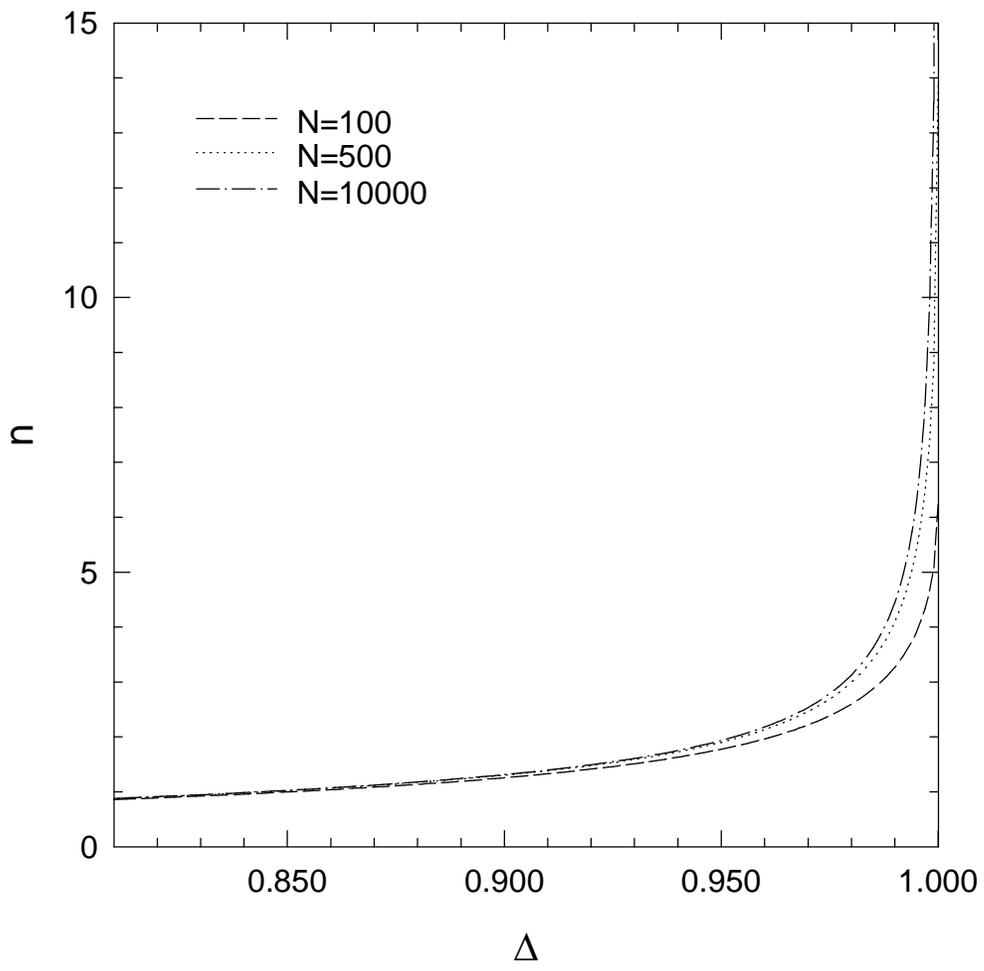
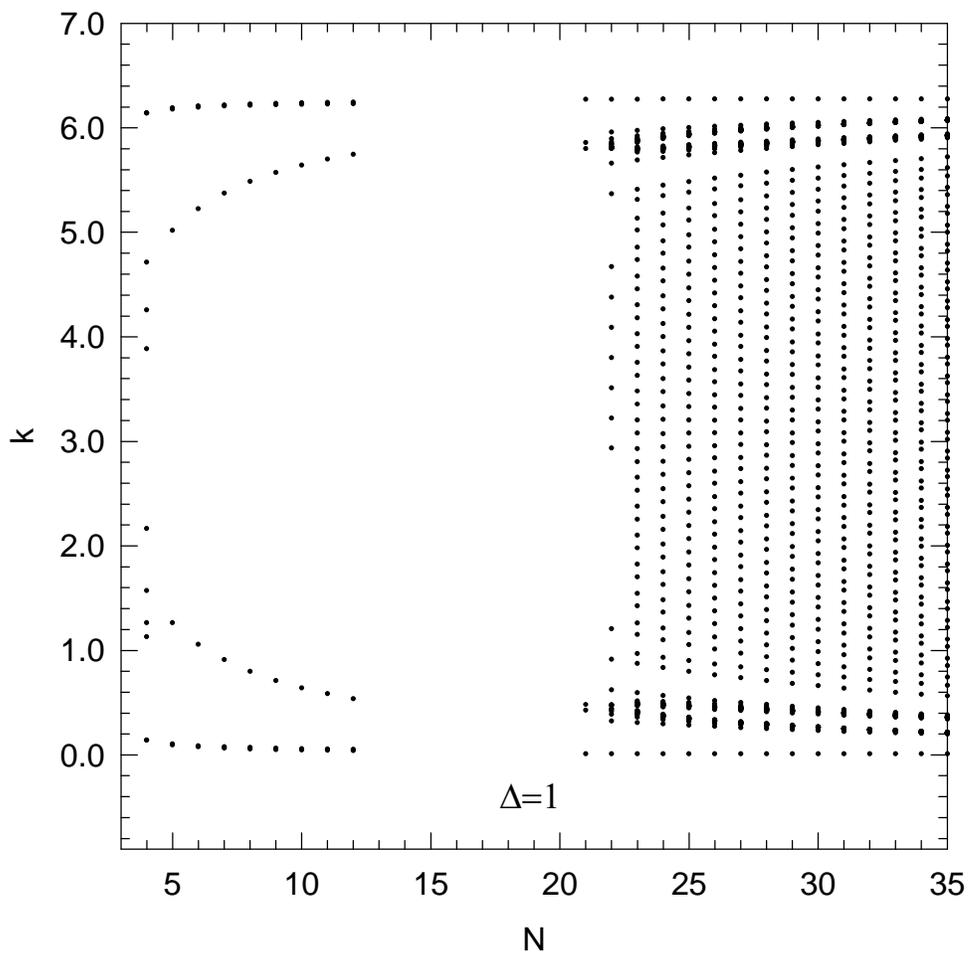
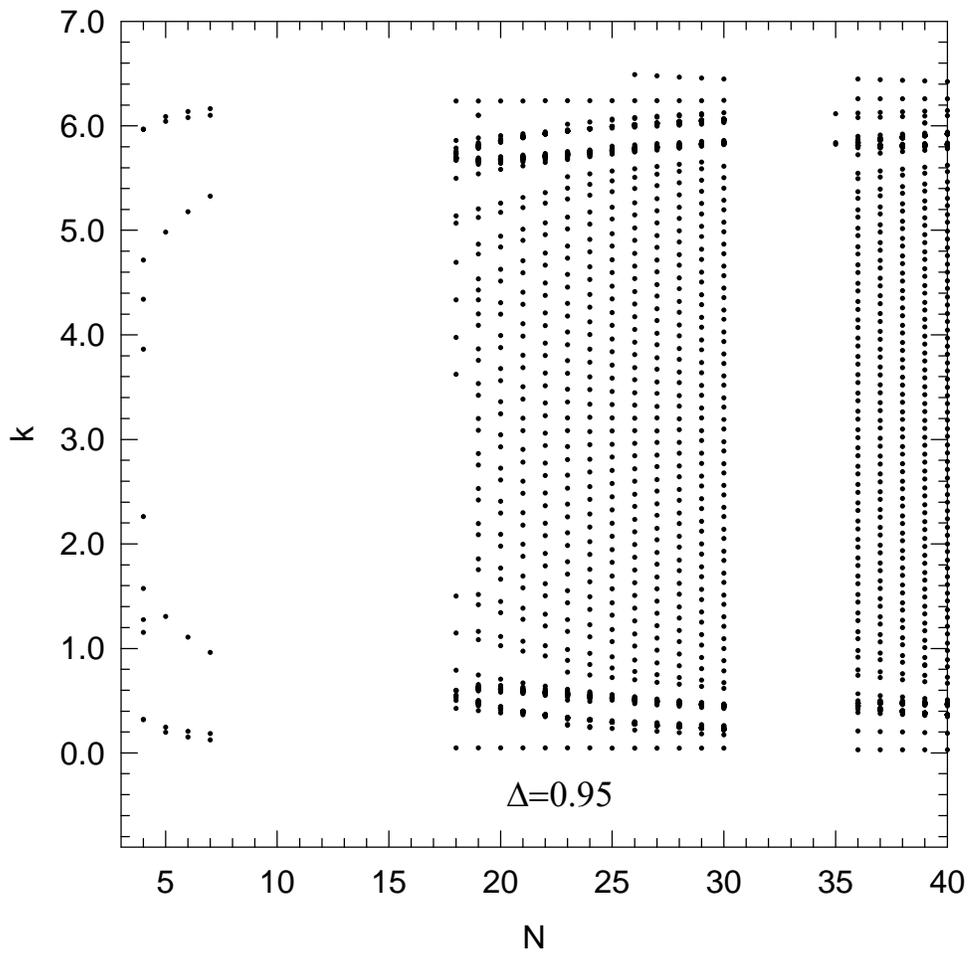
\end{document}